%% file: rxj0134.tex
\documentclass{aa}

\usepackage{epsf}
\usepackage{graphics}

\begin{document}

\input clipfig.tex
\useunitmm


\newcommand{\lb}{$\lambda$}
\newcommand{\sm}[1]{\footnotesize {#1}}
\newcommand{\inft}{$\infty$}
\newcommand{\vlv}{$\nu L_{\rm V}$}
\newcommand{\lv}{$L_{\rm V}$}
\newcommand{\lx}{$L_{\rm x}$}
\newcommand{\lsoft}{$L_{\rm 250eV}$}
\newcommand{\lhard}{$L_{\rm 1keV}$}
\newcommand{\vlsoft}{$\nu L_{\rm 250eV}$}
\newcommand{\vlhard}{$\nu L_{\rm 1keV}$}
\newcommand{\vlir}{$\nu L_{60\mu}$}
\newcommand{\ax}{$\alpha_{\rm x}$}
\newcommand{\aopt}{$\alpha_{\rm opt}$}
\newcommand{\aoxs}{$\alpha_{\rm oxs}$}
\newcommand{\aoxh}{$\alpha_{\rm oxh}$}
\newcommand{\airhard}{$\alpha_{\rm 60\mu-hard}$}
\newcommand{\aoxsoft}{$\alpha_{\rm ox-soft}$}
\newcommand{\aio}{$\alpha_{\rm io}$}
\newcommand{\aixs}{$\alpha_{\rm ixs}$}
\newcommand{\aixh}{$\alpha_{\rm ixh}$}
\newcommand{\hb}{H$\beta_{\rm b}$}
\newcommand{\nh}{$N_{\rm H}$}
\newcommand{\nhgal}{$N_{\rm H,gal}$}
\newcommand{\nhfit}{$N_{\rm H,fit}$}
\newcommand{\ale}{$\alpha_{\rm E}$}
\newcommand{\cts}{$\rm {cts\,s}^{-1}$}
\newcommand{\pl}{$\pm$}
\newcommand{\kev}{\rm keV}
\newcommand{\rb}[1]{\raisebox{1.5ex}[-1.5ex]{#1}}
\newcommand{\ten}[2]{#1\cdot 10^{#2}}
\newcommand{\msun}{$M_{\odot}$}
\newcommand{\dM}{\dot M}
\newcommand{\dMM}{$\dot{M}/M$}
\newcommand{\dMedd}{\dot M_{\rm Edd}}
\newcommand{\kms}{km\,$\rm s^{-1}$}

\thesaurus{03(02.02.1; 11.01.2; 11.14.1; 11.17.4 RX J0134.2--4258)}

\title{The enigmatic Soft X-ray AGN RX J0134.2--4258
\thanks{Based in part on
observations at the European Southern Observatory La Silla (Chile) 
with the 2.2m telescope of the Max-Planck-Society during MPG and ESO time, 
and with the ESO 1.5m telescope in September/October 1995}
}
\author{D. Grupe\inst{1}
\and K.M. Leighly\inst{2}
\and H.-C. Thomas\inst{3}
\and S.A. Laurent-Muehleisen\inst{4}
}
\offprints{\\ D. Grupe (dgrupe@xray.mpe.mpg.de)}
\institute{MPI f\"ur extraterrestrische Physik, Giessenbachstr., 
85748 Garching, FRG
\and Columbia Astrophysics Laboratory, 550 West 120th St., New York, 
NY 10027, U.S.A.
\and MPI f\"ur Astrophysik, Karl-Schwarzschildstr. 1, 85748 Garching, FRG
\and UC-Davis and IGPP/LLNL, 7000 East Av., Livermore, CA 94550, USA
}
\date{Received 08 June 1999/ Accepted 12 January 2000 }
\maketitle

\begin{abstract}
We report the discovery and analysis of the follow-up ROSAT pointed
observation, an ASCA observation and optical and radio observations
of the enigmatic soft X-ray AGN RX J0134.2--4258.  In the optical, RX
J0134.2--4258 appears as an extreme 'Narrow-Line Seyfert 1 galaxy'
(NLS1), with very strong FeII emission, very blue optical continuum
spectrum and almost no Narrow Line Region emission.  While its
spectrum was one of the softest observed from an AGN during the ROSAT
All-Sky Survey, its spectrum was found to be dramatically harder
during a pointed observation although the count rate remained
constant. We found in the pointed observation that the spectrum is
softer when it is fainter and spectral fitting demonstrates that it is
the hard component that is variable.  The ASCA observation confirms
the presence of a hard X-ray power law, the slope of which is rather
flat compared with other NLS1s.  Survey and followup radio
observations reveal that RX J0134.2--4258 is also unusual in that it is
a member of the rare class of radio-loud NLS1s, and, with R=71, it
holds the current record for largest radio-to-optical ratio in NLS1s.
We discuss possible scenarios to explain its strange behaviour.
\end{abstract}
\section{Introduction}
With the launch of the X-ray satellite ROSAT in 1990 (Tr\"umper 1983)
a new window was opened for astronomy. One of ROSAT's first tasks was
to perform the first all-sky survey in the 0.1-2.4 keV energy band,
the ROSAT All-Sky Survey (RASS; Voges et al. 1993, 1999).  A large
number of Active Galactic Nuclei (AGN) with extremely steep X-ray
spectra were discovered in this survey (e.g. Thomas et al.  1998,
Grupe et al. 1998a, 1999, Beuermann et al. 1999).    Some
soft X-ray AGN appear to be transient in the ROSAT Position Sensitive
Proportional Counter energy window (PSPC, Pfef\-fermann et al. 1986).
These AGN were bright only in their 'high' state during the RASS.
Some prominent examples are IC 3599 (Brandt et al. 1995, Grupe et
al. 1995a), WPVS007 (Grupe et al. 1995b), and NGC 5905 (Bade et al.
1996, Komossa \& Bade 1999).  Both IC 3599 and NGC 5095 showed a
similar decrease by about a factor 70 which suggests that they were
experiencing an X-ray outburst during the RASS.  WPVS007 showed a
decrease by a factor of $\sim$ 400 between the RASS and a pointed
observation about three years later, and was practically turned off
after that.  The situation in RX J0134.2--4258 appears to be
different. While its count rate remained nearly constant between the
RASS and the pointed PSPC observation about two years later, its
spectral shape changed dramatically.  Together with WPVS007, it had
the steepest soft X-ray spectrum during the RASS, but in the pointed
observation it showed a flat spectrum (Mannheim et al. 1996).
Follow-up medium resolution spectroscopy showed this object to be a
Narrow-Line Seyfert 1 galaxy with a blue optical spectrum and very
strong FeII emission. It is one of the most peculiar sources in our
sample of bright soft X-ray selected ROSAT AGN (see Grupe et
al. 1998a, 1999).

In Sect. \ref{observe} we describe the observations and data reduction.
Sect. \ref{results} gives a summary of the results of the ROSAT, ASCA and
optical observations, and in Sect. \ref{discussion} we discuss possible
scenarios that may explain the strange behaviour found in this object.

\section{\label{observe} Observations and data reduction}
RX J0134.2--4258 was observed during the RASS between 1990, December
09, 20:42h and 1990, December 17, 23:11h for a total of 574 s.
Photons were collected in a circle of 250'' in radius for the source
and two circles of 400'' in the scan direction for the background.

About two years after the RASS observation, RX J0134.2--4258 was
observed again by ROSAT between 1992, December 28, 19:08h and 1993,
January 03, 23:37h, for a total of 7467 s. The same size extraction
regions were used, with the background taken from a source-free area
near RX J0134.2--4258.  The pointed observation count rate was
measured in channels 11-201, and spectral fitting was performed over
the same band. 
All power law model index values are quoted in terms of
the energy spectral index $\alpha_{\rm X}$ with $\rm F_{\nu} \propto
\nu^{-\alpha}$.
The hardness ratio was defined
by $HR = \frac{hard - soft}{hard + soft}$, where the soft and hard
bands correspond to channels 11-41 and 52-201, respectively.
RX J0134.2--4258 was also observed with the ROSAT High Resolution Imager (HRI)
between 1996, June 17, 01:57h and 03:22h (UT) for a total of 2393 s.

Our ASCA (Tanaka et al. 1994) observation was performed on 1997
December 10.  Standard methods using FTOOLS 4.1 were used to reduce
the data.  The net exposures were 45.6, 45.8, 51.0, 51.0 ks for SIS0,
SIS1, Gas Imaging Spectrometer (GIS) 2 and GIS3, respectively.

Optical spectra were obtained at different times.  The first
low-resolution spectrum was obtained in August 1993 with the 2.2m
MPG/ESO telescope in La Silla. A medium resolution spectrum taken
about a month later under photometric conditions at the same telescope
is presented in Grupe et al. 1999. The spectrum we present in this
paper, which has the best signal to noise, was taken in four nights in
September/October 1995 using the Boller \& Chivens spectrograph at the ESO
1.5m telescope with grating \#23 which has 600 grooves $\rm mm^{-1}$
and a dispersion of 126 \AA~$\rm mm^{-1}$ in first order. The spectral
resolution was $\sim$ 5\AA.  The camera was equipped with a FA 2048L
2048$\times$2048 CCD with a pixel size of 15$\times$15$\rm \mu m^{2}$
(ESO CCD \#24). The total observing time was 3.25 hours. The
conditions were cloudy and therefore the absolute flux cannot be
measured from this spectrum.  We measure $\rm m_V=16.0$  from
the spectrum obtained under photometric conditions (Grupe et al. 1999).
RX J0134.2--4258 was also observed by the HST (Goodrich et al. in prep.)
From this spectrum $\rm m_V=16.2$ was measured.

A radio source was detected in the Parkes-MIT-NRAO (PMN) Surveys near
the position of RX~J0134.2--4258 (Wright et al.\ 1994).  
In order to obtain a better radio position, on May 7 1999 we observed
RX J0134.2--4258 at 8.4 GHz while the VLA\footnote{The VLA and NRAO are
operated by Associated Universities, Inc., under cooperative agreement with
the National Science Foundation.} was in the D-configuration.  Time on source
was approximately 36 min.  Absolute flux calibration was set by a short
observation of 3C\,286 and another calibrator close in position to
RX J0134.2--4258 was observed to correct the phases.  Data were calibrated and
reduced in the standard iterative manner using the NRAO's AIPS analysis
package (v15APR98) and using the tasks MX and CALIB.  Self-calibration for
correction of only the phases was applied in order to preserve the absolute
flux density scale.  The noise in the final map is 0.18 mJy and because of the
very low declination of this source, the beam is an elongated $41.6\arcsec
\times 5.2\arcsec$ at an PA of $\approx 0^{\circ}$.  The source is
completely unresolved in our map.

\section{\label{results} Results}
\begin{figure}[t]
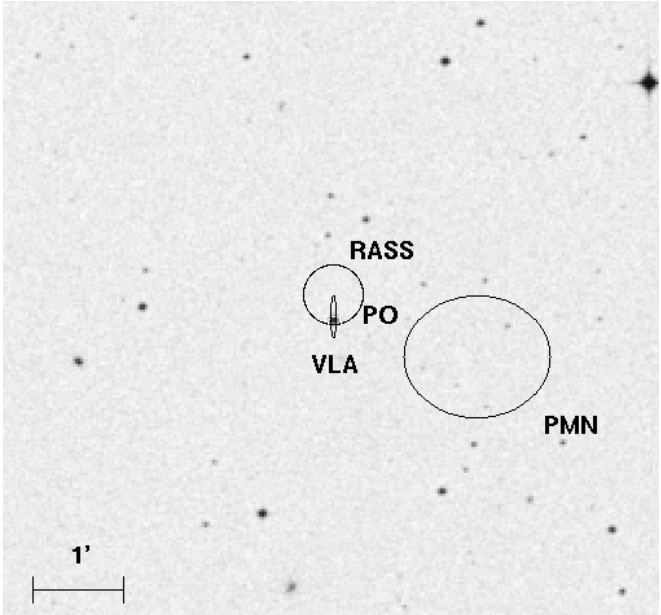

\clipfig{such_rxj0134}{87}{32}{107}{185}{251}
\caption[ ]{\label{find}
Finding chart of RX J0134.2--4258. The circles/ellipes mark the
$2\sigma$ error of the RASS, the ROSAT pointed observations (PO), the
PMN radio survey, and the VLA observation.  See text for details. The
ROSAT pointed observation error circle fell directly on the source.
}
\end{figure}

Fig. \ref{find} shows the $7'\times 7'$ finding chart of RX
J0134.2--4258 and the measured X-ray, optical, and radio positions of
RX J0134.2--4258 are given in Table \ref{position}.  The 2$\sigma$
(90\% confidence) error circles of the RASS and pointed observations
are marked. The optical position was derived from the US Naval
Observatory scans of the ESO/SRC plates.  The X-ray source corresponds
to an optical counterpart that was identified as an AGN at z=0.237
(Grupe et al. 1998a).  The error circle of the RASS observation was
estimated by fitting a Gaussian profile to the data and averaging the
positional uncertainties in RA and DEC. The position and error circle
of the pointed observation was determined by using a maximum
likelihood algorithm in the EXSAS software package (Zimmermann et al.\
1998).

The 4.85~GHz radio position was obtained in the Parkes-MIT-NRAO (PMN)
radio survey (Wright et al. 1994).  The uncertainty on this position,
which depends on the location and flux of the object (see Wright et
al. 1994 for details), indicate that the optical position of RX
J0134.2--4258 is nearly $3\sigma$ in RA from the cataloged radio
source (Fig. \ref{find}).  This fact, casting significant doubt on
the possibility that the optical and radio sources were associated,
prompted us to make a new pointed observation using the VLA.  We used
the task JMFIT on the new VLA data to fit a gaussian to the peak of
the emission.  The position and uncertainty (taken to be 1/3 the beam
in each direction) are shown on Fig. \ref{find} and listed in Table
\ref{position}.  We assert that the coincidence of the radio source
and RX J0134.2--4258 is such to warrant a confident association of the
optical/X-ray source with this radio object.

\begin{figure}[h]
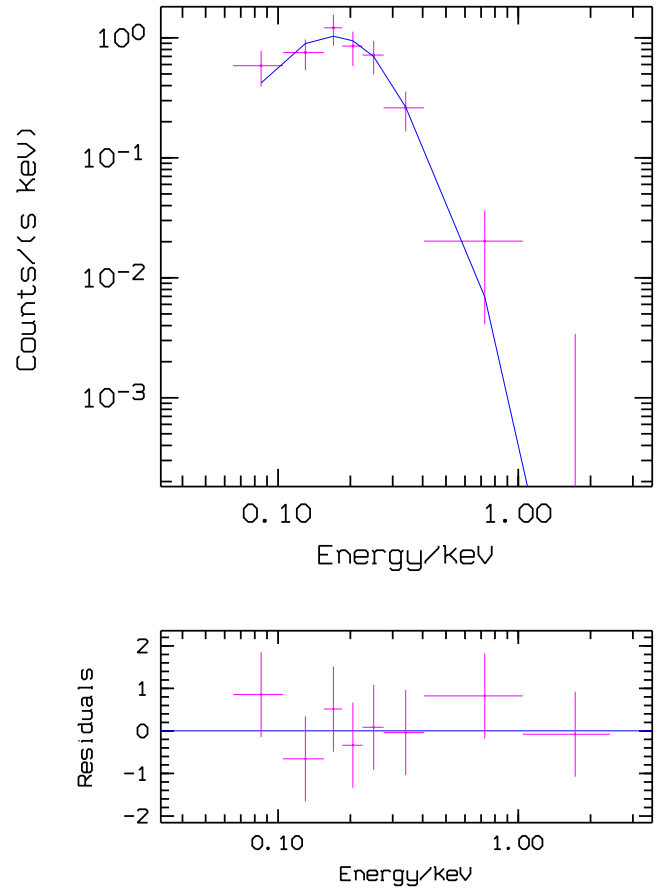

\clipfig{rxj0134_rass}{87}{18}{16}{140}{185}
\caption[ ]{\label{rass_spec}
Power law fit to the RASS spectrum of RX J0134.2--4258. The cold absorption
parameter was fixed to the Galactic value. The X-ray spectral slope is 
\ax~=~6.9\pl2.9.
}
\end{figure}

\begin{table*}
\caption{\label{position} X-ray, optical, and radio
positions of RX J0134.2--4258.
``$2\sigma$'' marks the $2\sigma$ error radius, and $\Delta$ markes the 
distances of the radio and X-ray positions to the optical position, 
 all given in arcsec}
\begin{flushleft}
\begin{tabular}{lllccc}
\hline\noalign{\smallskip}
Observation & $\alpha_{2000}$ & $\delta_{2000}$ & $\alpha~2\sigma$ & 
$\delta~2\sigma$ & $\Delta$ \\
\noalign{\smallskip}\hline\noalign{\smallskip}
RASS & 01 34 16.9 & --42 58 08 & 20.0 & 20.0 & 20 \\
Pointed Obs. & 01 34 16.7 & --42 58 28 & 2.0 & 2.0 & 2 \\
Optical(USNO) & 01 34 16.9 & --42 58 27 & 0.5 & 0.5 & 0.0 \\
Radio (PMN 4.85 GHz) & 01 34 10.6 & --42 58 49 & 48 & 40 & 55 \\
Radio (VLA 8.4 GHz) & 01 34 16.9 & --42 58 22 & 1.7 & 14  & 5 \\
\noalign{\smallskip}\hline\noalign{\smallskip}
\end{tabular}
\end{flushleft}
\end{table*}

\subsection{ROSAT}

The RASS observation yielded 120\pl29 counts, giving a count rate of
0.21\pl0.05 $\rm cts~s^{-1}$ and an HR of $-$0.84\pl0.05 (Grupe et
al. 1998a, 1999).  A power law plus absorption model reveals that the
intrinsic absorption in this source appears to be negligible and
therefore the absorption was fixed at the Galactic value ($N_{\rm
H}~=~1.59~10^{20}~\rm cm^{-2}$; Dickey \& Lockman 1990).  The
resulting energy spectral index is very steep: \ax~=6.9\pl2.9.  Spectral fitting
results are given in Table \ref{fits}, and the result of this fit is
shown in Fig. \ref{rass_spec}.  A blackbody model was also
fitted. The absorption column density
approached zero when it was allowed
to be free and therefore it was again fixed at the Galactic value.
The resulting radiation temperature was 24\pl8 eV. We cannot
distinguish between a power law and a black body model for the RASS
data.  The low signal to noise of the spectrum did not allow more
sophisticated fits.

The pointed ROSAT PSPC observation yielded 1476\pl43 counts corresponding to a
count rate of 0.20\pl0.01 $\rm cts~s^{-1}$.  This was nearly the same
rate as was found in the RASS observation; however, the spectrum was
much harder (HR=$-$0.12\pl0.01).  A single power law fit to the total
pointed observation spectrum with column density fixed to the Galactic
value does not give a satisfactory description of the soft spectrum
(Fig. \ref{po_spec}, left panel).  A single black body model does
not give a good fit either. A broken power law gave a much better fit
(Fig. \ref{po_spec}, middle panel).  However, to determine the
uncertainties on the model parameters, it was necessary to fix the
slope of the hard power law.  We used two values: \ax=1.0, the result
of the free fit, and \ax=0.8, the result from the ASCA spectra (see
Sect. \ref{asca_data}).  Interestingly, for both cases, we found
that the soft X-ray component had a similar slope as was found during
the RASS observation.  A black body plus a power law also fit the
spectrum well (Fig. \ref{po_spec}, right panel); in this case, all
parameters except for the absorption, which was still fixed to the
Galactic value, could be left free.  These spectral fitting results
are also summarized in Table \ref{fits}.

We can infer the presence of a soft excess if a power law plus
absorption model yields a column density less than the Galactic value.
Fig. \ref{fit_grid} displays the contours of power law fits to the
RASS and pointed observation spectra with the absorption column free.
The RASS spectra are consistent with a single power law model;
however, the measured column density
for the pointed observation is less than
the Galactic column density, indicating the presence of a soft excess.

\def \charthoffset  {\hspace{0.2cm}}
\def \charthsep     {\hspace{0.3cm}}
\def \chartvsep     {\vspace{0.3cm}}
\newcommand{\putcharta}[1]{\clipfig{#1}{57}{18}{16}{140}{185}}
\newcommand{\chartlinea}[3]{\parbox[t]{18cm}{

\noindent\charthoffset\putcharta{#1}\charthsep\putcharta{#2}\charthsep\putcharta{#3}}}

\newcommand{\putchartb}[1]{\clipfig{#1}{80}{24}{10}{215}{195}}
\newcommand{\chartlineb}[2]{\parbox[t]{18cm}{

\noindent\charthoffset\charthoffset\putchartb{#1}
\charthsep\charthsep\charthsep\charthsep\putchartb{#2}}}

\begin{figure*}[t]
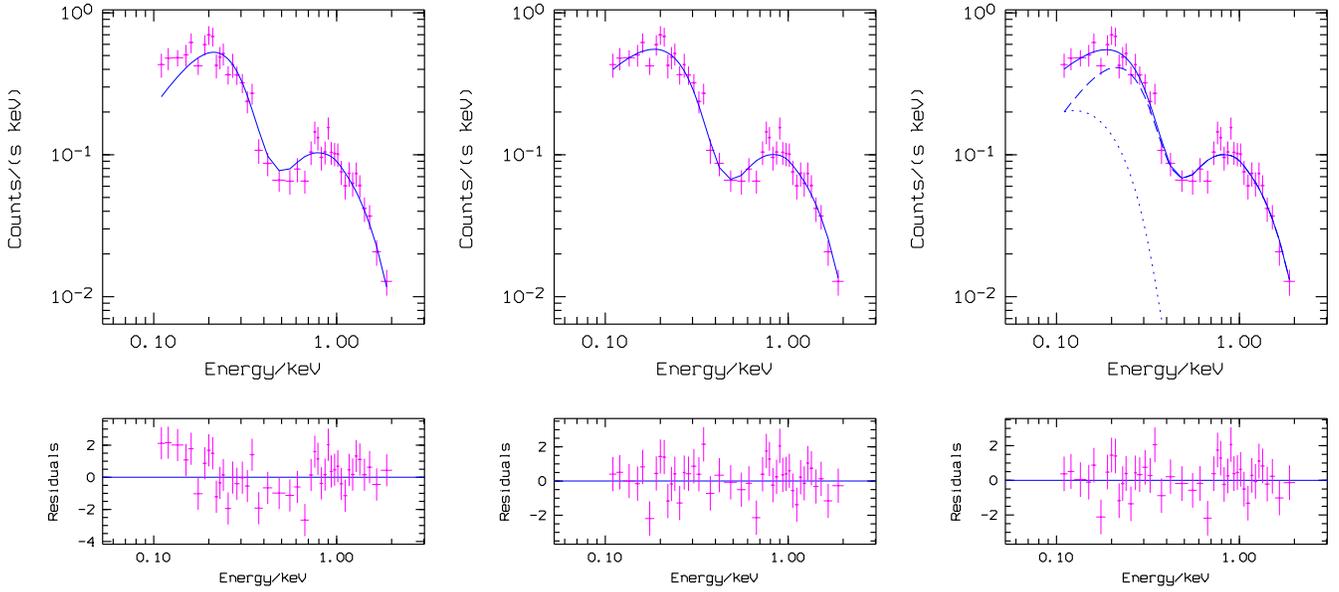

\chartlinea{rxj0134_po_pl}{rxj0134_po_bp}{rxj0134_po_bbpl}
\caption[ ]{\label{po_spec}
Spectral fits to the pointed observation 
spectrum of RX J0134.2--4258: a single
power law fit (left panel), 
with $N_{\rm H}$ fixed to the Galactic value;
a broken power law fit(middle panel), 
also fixed $N_{\rm H}$  and the high energy
power law slope fixed to \ax=1.0; and an absorbed
black body plus power law spectral fit (right panel). 
The fit parameters are given in Table \ref{fits}.
}
\end{figure*}

\begin{figure*}[t]
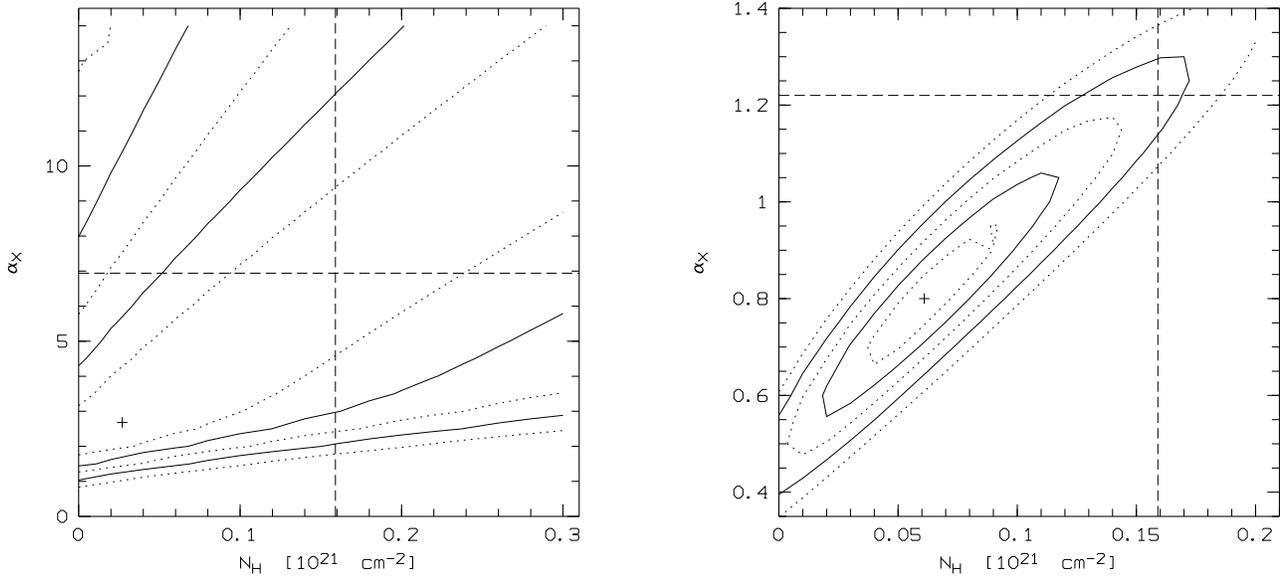

\chartlineb{rxj0134_rass_pl_grid}{rxj0134_po_pl_grid}
\caption[ ]{\label{fit_grid}
Contour plots of power law fits to the X-ray data. The left panel shows the
RASS data and the right one the pointed observation.  The dashed straight lines
mark the results of a fit with fixed column density ($N_{\rm H}$ =0.159).
The contours are at 1, 2, 3, 4, and 5 $\sigma$ level.
}
\end{figure*}

\begin{table*}
\caption{\label{fits} Spectral fits to the RASS and pointed observation X-ray
data of RX J0134.2--4258. ``$N_{\rm H}$'' is the column density given in 
units of $10^{20} \rm cm^{-2}$, ``Norm'' is the normalization at 250 eV (rest
frame) in $\rm 10^{-3}~Photons~cm^{-2}~s^{-1}~keV^{-1}$, 
``$\alpha_{\rm X-soft}$'' the soft energy spectral slope, ``Break'' the 
break energy in keV of the broken power law fit (rest frame), 
``$\alpha_{\rm X-hard}$'' the hard energy spectral slope, ``A'' the black body
integral in $Photons~cm^{-2}~s^{-1}$, and ``$T_{\rm rad}$'' the radiation
temperature in eV.
The models used here 
are simple power laws (powl), broken power laws (brpl), Black body (bbdy),
and a black body plus power law (bbpl).
}
\begin{flushleft}
\begin{tabular}{llcccccccc}
\hline\noalign{\smallskip}
Observation & model & $N_{\rm H}$ & Norm &  $\alpha_{\rm X-soft}$ &
Break & $\alpha_{\rm X-hard}$ & A & $T_{\rm rad}$ & $\chi^{2}/\nu$
\\
\noalign{\smallskip}\hline\noalign{\smallskip}
RASS & powl & 0.27\pl1.11 & 7.20\pl1.95 & 2.68\pl5.63 & 
--- & --- & --- & --- & 1.5/5 \\
RASS & powl & 1.59 fix & 6.04\pl4.97 & 6.94\pl2.60 & --- & ---
& --- & --- & 2.2/5 \\ 
RASS & bbdy & 1.59 fix & --- & --- & --- & --- & 
0.63\pl1.45 & 24\pl8 & 2.5/4 \\
RASS & bbpl & 1.59 fix & 0.43\pl1.44 & --- & --- & 1.00 fix & 
0.72\pl0.15 & 23 fix & 2.3/6 \\
Pointed  & powl & 0.61\pl0.37 & 5.49\pl1.57 & 0.80\pl0.07 & --- & ---
& --- & --- & 43/40 \\
Pointed & powl & 1.59 fix & 10.20\pl0.56 & 1.22\pl0.06 & --- & ---
& --- & --- & 58/41 \\
Pointed & brpl & 1.59 fix & 12.90\pl10.50 & 6.17\pl5.00 & 0.27\pl0.02 &
1.00 fix & --- & --- & 40/40 \\
Pointed & brpl & 1.59 fix & 17.11\pl1.40 & 3.83\pl1.46 & 0.34\pl0.05 &
0.80 fix & --- & --- & 44/40 \\ 
Pointed & bbdy & 1.59 fix & --- & --- & --- & --- & 0.004\pl0.0003 & 145\pl4
& 349/41 \\
Pointed & bbpl & 1.59 fix & 10.0\pl9.5 & --- & --- & 1.04\pl1.22 & 8.75\pl0.01
& 14\pl7 & 40/39 \\
Pointed & bbpl & 1.59 fix & 14.6\pl3.3 & --- & --- & 1.00 fix & 
0.15\pl0.03 & 23 fix & 42/41 \\
OBI 3 & bbpl & 1.59 fix & 2.25\pl1.58 & --- & --- & 1.00 fix & 0.20\pl0.89 
& 23\pl3 & 8.5/8 \\
OBI 5 & bbpl & 1.59 fix & 15.3\pl17.6 & --- & --- & 1.00 fix & 0.12\pl1.5 
& 23\pl32 & 7.3/11 \\
\noalign{\smallskip}\hline\noalign{\smallskip}
\end{tabular}
\end{flushleft}
\end{table*}

\begin{figure}[t]
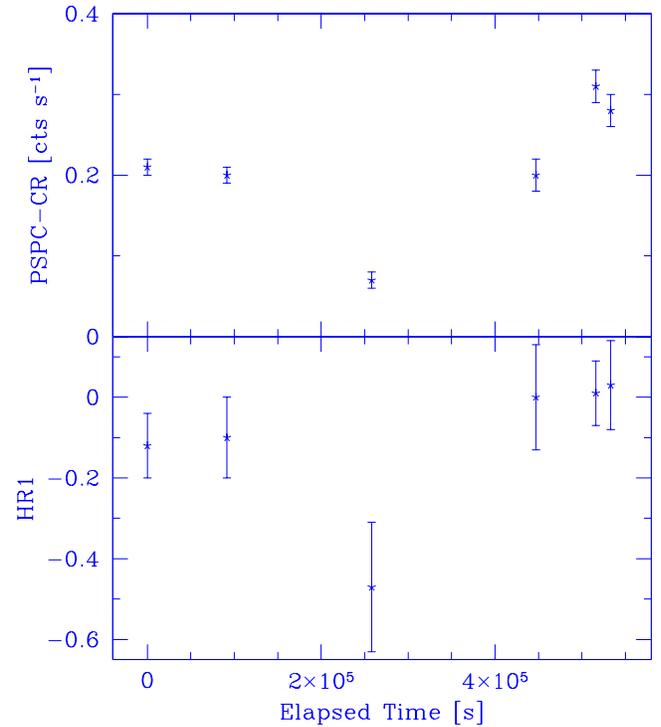

\clipfig{rxj0134_obis}{87}{12}{72}{162}{240}
\caption[ ]{\label{light}
Light curve of RX J01334.2--4258 during the pointed observation between
92/12/28 and 93/01/03 (OBIs 1-6). The start time is 92/12/28 19:08 (UT).
The upper panel shows the change in the 
count rate and the lower one the spectral behaviour in the hardness ratio 1.
}
\end{figure}

\begin{figure}[t]
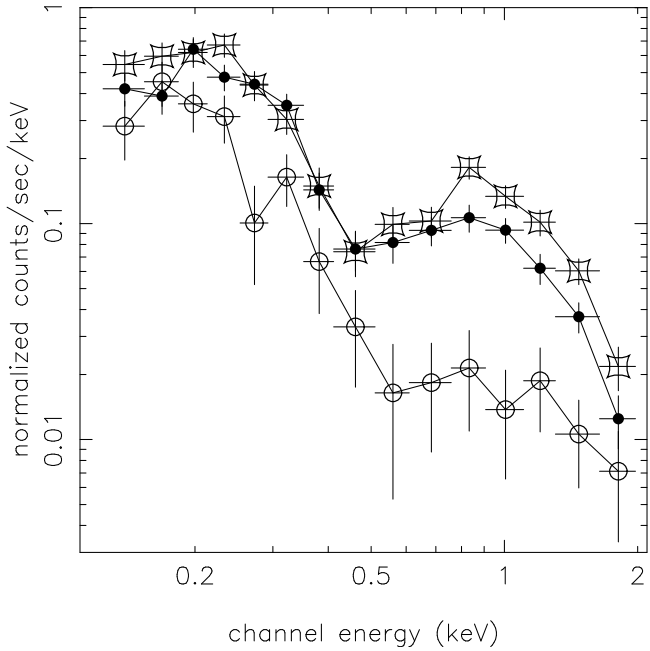

\clipfig{rxj0134_obi_spec}{87}{20}{50}{180}{210}
\caption[ ]{\label{light_spec}
Spectra of the ROSAT pointed observation. 
The solid circle are OBIs 1 and 2, the open
circles are OBI 3, and the squares are OBIs 4-6.
}
\end{figure}

No significant variability was detected from the RASS light curve;
however, the statistics are poor enough that we cannot exclude
variability with high confidence.  The pointed observation consisted
of 6 observation intervals (OBI); each OBI corresponds roughly to a 1
or 2 orbit exposure.  Binning the data in each OBI showed clear
variability in both the flux and hardness ratio (Fig. \ref{light}).
The spectral variability is correlated with the flux variability in
that the spectrum becomes {\it harder} when the count rate {\it
increases}. Fig. \ref{light_spec} displays how the spectra
changed. While the soft channels remain practically constant, the hard
channels became fainter by a factor of almost 10.  To investigate this
further, we made spectral fits of OBIs 3 and 5 using a two component
spectral model consisting of an absorbed black body and a power law
(Table \ref{fits}). These fits show that the radiation temperature of
the spectra remains the same, and the biggest variation is seen in the
normalization of the power law component.  This means that the
spectral variability can be consistently explained as originating from
variability in the flux of the hard component.
 
RX J0134.2--4258 was not detected during the ROSAT HRI observation.
An upper limit of 0.0011 cts $\rm s^{-1}$ was measured. Assuming the
power law fit to the OBI 3 data, a count rate of 0.0028 cts $\rm
s^{-1}$ was expected.  We infer that during the one and a half hours
of the observation the hard X-ray component was in its low state.  The
HRI is much less sensitive to lower energies than the PSPC; therefore
the soft X-ray component was not detected.

\subsection{\label{asca_data} ASCA}

\begin{figure}[t]
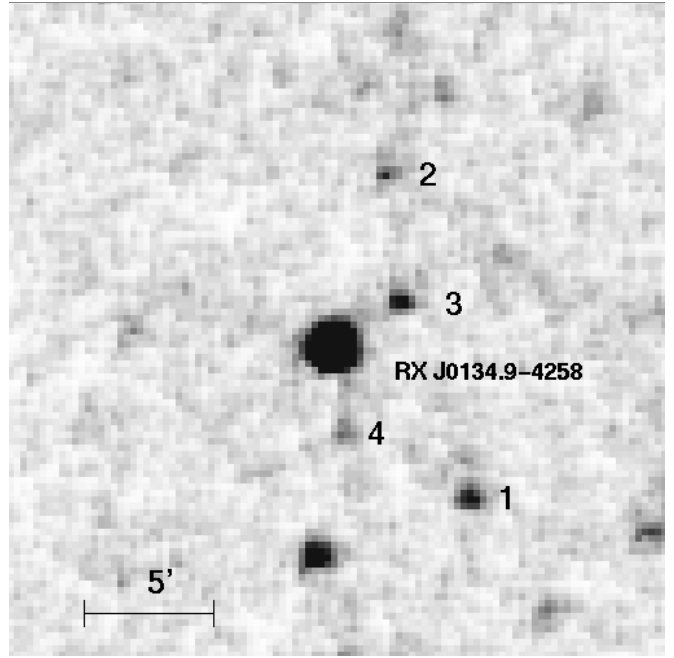

\clipfig{rxj0134_pspc_image}{87}{32}{108}{185}{261}
\caption[ ]{\label{rosat_image} ROSAT PSPC image from the pointed
observation of RX J0134.2--4258. The positions of the surrounding
sources 1 - 4 are given in Table \ref{position_list}.  }
\end{figure}

\begin{figure}[t]
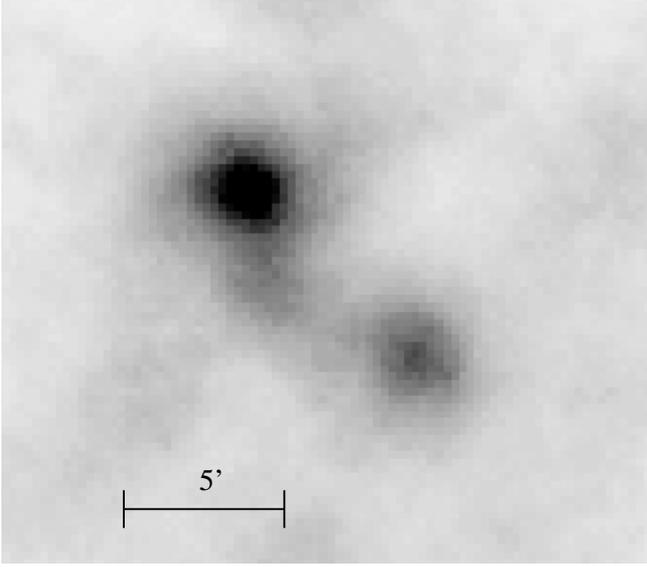

\clipfig{rxj0134_asca_image}{87}{54}{98}{164}{193}
\caption[ ]{\label{asca_image}
ASCA 2-5 keV GIS image of RX J0134.2--4258, the brightest source in the image. 
The weak
source about 3.5' south of RX J0134.2--4258 corresponds to source \#4 in Figure
\ref{rosat_image} and the other brighter source about 7' southwest of RX
J0134.2--4258 corresponds to source \#1.
}
\end{figure}

RX~J0134.2--4258 was clearly detected in our ASCA images. Figure
\ref{asca_image} shows the smoothed, combined GIS2 and GIS3 {\it ASCA}
image in the hard (2.0--5.0 keV) band; RX J0134.2--4258 is the
brightest source in the field, but several other sources are visible.
The superior Point Spread Function (PSF) available from the {\it
ROSAT} PSPC allowed us to identify several nearby X-ray emitting
sources that could contaminate the {\it ASCA} spectrum (Figure
\ref{rosat_image}) and their positions are listed in Table
\ref{position_list}.  None of these objects have catalogued
identifications.

\begin{table}
\caption{\label{position_list} X-ray positions of the sources
surrounding RX J0134.2--4258.  The source number corresponds to the
marked positions in Fig. \ref{rosat_image}.  }
\begin{flushleft}
\begin{tabular}{lll}
\hline\noalign{\smallskip} Source & $\alpha_{2000}$ & $\delta_{2000}$
\\ \noalign{\smallskip}\hline\noalign{\smallskip} \#1 & 01 33 20.5 &
--43 04 31.5 \\ \#2 & 01 34 04.5 & --42 51 34.7 \\ \#3 & 01 34 01.9 &
--42 56 38.2 \\ \#4 & 01 34 12.7 & --43 01 34.8 \\
\noalign{\smallskip}\hline\noalign{\smallskip}
\end{tabular}
\end{flushleft}
\end{table}

The nominal source extraction regions, with radii 4 and 6 arcminutes
for the SIS and GIS respectively, were used initially.  The background
spectra were estimated using source-free regions of the detectors.
The source was not detected at high energies and thus the spectra were
fitted in the energy bands 0.5--7.0 and 0.8--8.0 keV for the SIS and
GIS respectively.  The net count rates were 0.017, 0.016, 0.010 and
0.013 $\rm cts~s^{-1}$ in SIS0, SIS1, GIS2 and GIS3, respectively. We
fitted all four spectra separately, and found that, as usual, SIS1
yielded slightly flatter spectral parameters and higher absorption.
However, the four detectors were consistent within 90\% confidence and
therefore all data from all detectors were used for spectral fitting.

Spectral fitting revealed evidence that the {\it ASCA} spectrum is
contaminated by a nearby source.  For the source extraction regions
listed above, the GIS flux is about 40\% larger than that of the SIS.
While the cross calibration between detectors is not perfect, the
differences should not be more than about 10\% (this number is not
known exactly because of changes in the SIS efficiency).  The
suspected contaminating source is \#4 for the following reasons.
First, it is about $3.5'$ from RX~J0134.2--4258, and therefore it
falls within the $6'$ GIS extraction region.  Also, it can be seen as
a faint extension toward the south in the {\it ASCA} image (Figure
\ref{asca_image}); in contrast, source \#3, which is also about
$3.25'$ from RX~J0134.2--4258, is not clearly present in the ASCA
image.  Thirdly, the difference in fluxes between SIS and GIS
increases as the source extraction region sizes are increased.  This
is expected because the GIS has a broader point spread function, but
more importantly because the position of this contaminating source
means that it falls off the chip completely for SIS1 and partially for
SIS0.  The photon index from a power law fit to all four detectors
increases slightly as the source extraction region is decreased.  This
result suggests that the contaminating source may be somewhat harder
than RX~J0134.2--4258.  This source is less than 6\% as bright as
RX~J0134.2--4258 in the pointed ROSAT observation; however, the flux
measurements listed above suggest that it contributes more than $\sim
20$\% to the {\it ASCA} spectrum.  The difference may be due to a
harder spectrum for the contaminating source (although analysis of the
ROSAT spectrum does not clearly indicate this) or variability.  A
faint (magnitude $\sim 19$) counterpart is visible in the DSS image at
the X-ray position; however, there is no cataloged identification.
Because of the possible contamination, we list spectral fitting
results from the regions listed above, as well as SIS alone and
SIS+GIS with extraction regions $2'$ in radius at which the
contamination by a source $3.5'$ distant should be minimal.

\subsubsection{ASCA spectral fitting results}

\begin{figure}[t]
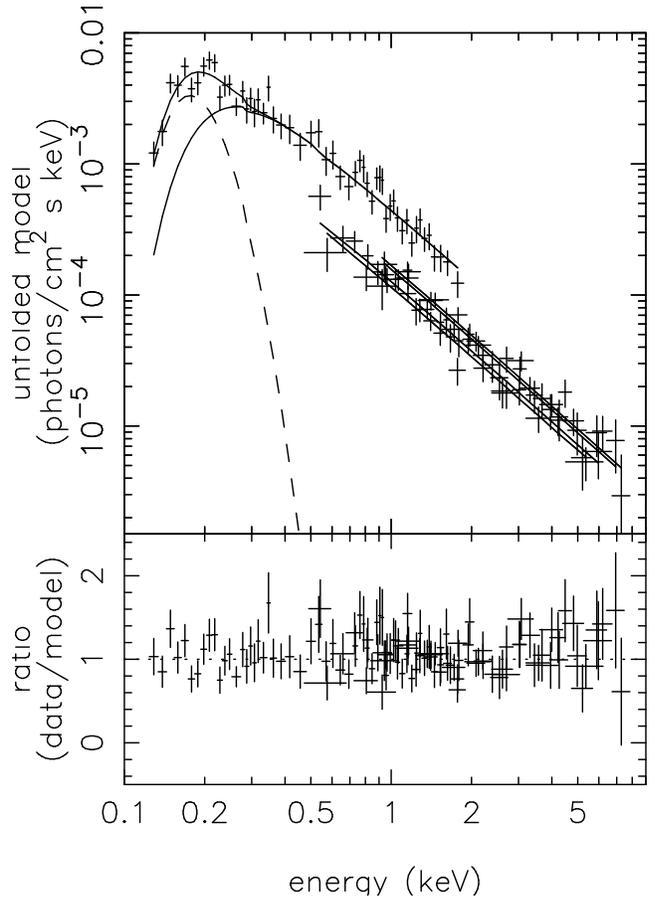

\clipfig{rxj0134_rosat_asca}{87}{6}{17}{180}{260}
\caption[ ]{\label{rosat_asca_spec}
Combined power law  + blackbody fit to the ROSAT and ASCA spectra.
}
\end{figure}

The spectra were first fitted with a power law plus Galactic
absorption, and a good fit was obtained (Figure
\ref{rosat_asca_spec}).  No features are apparent in the spectra.
Iron lines are frequently found in the {\it ASCA} spectra of Seyfert
galaxies, although there appears to be some dependence on luminosity
(e.g. Nandra et al. 1997).  Addition of a narrow ($\sigma=50\rm\,eV$)
iron line does not improve the fit; the upper limit is given in Table
\ref{asca_spec}.  Because the upper limit on the equivalent width is
consistent with those measured from iron lines detected in other
Seyfert galaxies as luminous as RX J0134.2--4258 ($1.4 \times 10^{37}$
W corresponding to a 2--10 keV flux of $4.8 \times 10^{-16}\rm
\,W\,m^{-2}$) we conclude that the statistics are too poor to
recognize the iron line if present.  Evidence for ionized or ``warm''
absorption is also frequently found in the {\it ASCA} spectra of
Seyfert galaxies (e.g. Reynolds 1997).  Addition of absorption edges
at 0.74 and 0.87~keV in the rest frame, relevant for absorption by
O~VII and O~VIII respectively, did not improve the fit significantly,
except in the case where the extraction region was $2'$
($\Delta\chi^2=5.1$ for 1 degree of freedom).  However, the observed
energy of this edge for z=0.237 is 0.6~keV, very near the edge of the
bandpass in the SIS in the observer's frame.  Since it is not detected
with the other extraction regions, the reality of this feature is
suspect.  The upper limits (Table \ref{asca_spec}) encompass typical
optical depths found in Seyferts and NLS1s (e.g.  Leighly 1999a) and we
conclude that the poor statistics prevent detection of a typical warm
absorber.  However, a warm absorber with very large optical depth
($\tau>1$) is clearly ruled out.

The power law energy spectral index in RX J0134.2--4258, measured to be 0.8,
has the distinction of being rather flat for an NLS1.  In a study of
25 spectra from 23 NLS1s, Leighly (1999a) found an average index \ax~
of 1.19 with an intrinsic dispersion of 0.3.  Therefore, such a flat
index is not unknown.  However, the other cases of best fit 
indices \ax~$<1.0$ occurred in Mrk~766 and NGC~4051, both of which have
somewhat peculiar properties for NLS1s, or in objects with complex
spectra and very weak hard tails (PHL~1092, PG~1404+226, and Mrk~507).
Steeper hard spectra were found in other objects.  Leighly (1999a) also
found that many NLS1s that have little evidence for absorption have
evidence for a soft excess.  There is a little evidence for a weak
soft excess in RX J0134.2--4258 in that that best fit value above
spectral slope above 2 keV, measured to be 0.6, is flatter than the
best fit index over the whole energy band.  However, the indices are
consistent within their statistical error, and addition of a soft
excess component does not improve the fit significantly.  It is
interesting to note, for reasons that will become apparent below,
that the spectral index is quite close to the average found from
radio-loud quasars observed by {\it Ginga}  ($0.72 \pm 0.07$; Lawson
\& Turner 1997). 

Because it appears that the ROSAT and ASCA spectral fitting results
may be consistent except for a normalization factor arising from
variability between the two observations, we proceed to
make a joint fit (Fig. \ref{rosat_asca_spec}, Table \ref{asca_spec}).
A power law with Galactic absorption provides an
acceptable but not a good fit ($\chi^2=309$ for 245 d.o.f.) and there
are residuals at low and high energies.  Addition of a black body improves the fit greatly
($\Delta\chi^2=43$ for two additional degrees of freedom).  
This joint fit shows that the power law
component normalization is a factor of about 3.5 larger in the {\it
ROSAT} pointed observation than in the {\it ASCA} observation.  Again,
no additional absorption, warm absorber as modeled by O~VII and O~VIII
edges or iron line are required.
It has to be noted that in one simultaneous ROSAT and ASCA observation
of NGC 5548, ROSAT data systematically appear steeper than ASCA data
(Iwasawa et al.  1999) and therefore there is some doubt whether the
cross calibration between the two missions is good enough to warrant a
simultaneous fit.  However, in this case, we see no evidence for a
cross calibration problem.

\begin{table}
\caption{\label{asca_spec} Spectral analysis of the ASCA observation plus
the combined ASCA + ROSAT spectral fit (below) 
}
\begin{flushleft}
\begin{tabular}{lccc}
\hline\noalign{\smallskip}
Parameter & All$^1$ & SIS only & 2' extr. reg.$^2$ \\ 
\noalign{\smallskip}\hline\noalign{\smallskip}
$\alpha_{\rm X}$  & $0.78^{+0.1}_{-0.09}$ & 0.77$\pm$0.12 &  0.85$\pm$0.09 \\
Flux SIS0  &    4.8       &      4.7       &      4.0 \\
Flux SIS1 (2-10 keV) &   4.8           &  4.8           &  4.2 \\
Flux GIS2            &   6.6           &  ---           &  4.5 \\
Flux GIS3            &   7.3           &  ---           &  4.5 \\
$\alpha_{\rm X}$ above 2keV     &   $0.62^{+0.26}_{-0.25}$ & 
                      $0.4_{-0.4}^{+0.9}$  &  $0.78^{+0.30}_{-0.29}$ \\
Fe flux limit  &  $<$2.3           & $<$4.7      &  $<$1.8 \\
Fe eqw limit (eV)     &  $<$260           & $<$550      &  $<$250 \\
0.74 keV edge depth   &  $<$0.61          & $<$0.74     &  
                       $0.83^{+0.53}_{-0.64}$ \\
0.87 keV edge depth   &  $<$0.43          & $<$0.46     &  $<$0.30 \\
Intrinsic abs (10$^{20}$) &  $<$0.4           & $<$0.5      &  $<$1.4 \\
\noalign{\smallskip}\hline\noalign{\smallskip}
\end{tabular}
\end{flushleft}

$^1$ SIS, GIS extraction regions 4 and 6 arcmin; may include
contaminating source.

$^2$ SIS, GIS extraction regions 2 arcmin

\begin{flushleft}
\begin{tabular}{lc}
\hline\noalign{\smallskip}
Parameter & ASCA + ROSAT fit \\ 
\noalign{\smallskip}\hline\noalign{\smallskip}
$\alpha_{\rm X}$ & 0.80$\pm$0.09 \\
kT (eV)      & $30^{+15}_{-9}$ \\
2--10 KeV flux (SIS0)$^1$ & 4.58 \\
2--10 keV luminosity$^2$ & 1.32 \\
0.1--2.0 keV flux$^1$ & 14.0 \\
0.1--2.0 keV luminosity$^2$ & 3.8 \\
neutral absorption & $<2.0~10^{20} \rm cm^{-2}$ \\
$\tau$ (0.74 keV) & $<$ 0.33 \\
$\tau$ (0.87 keV) & $<$ 0.27 \\
Fe K line flux$^3$ & $<$ 2.5 \\
Fe K equivalent width & $<$ 295 \\
\noalign{\smallskip}\hline\noalign{\smallskip}
\end{tabular}
\end{flushleft}

(Notes: confidence intervals are 90\% for one parameter of interest; i.e.
$\Delta\chi^2=2.71$.) \\
($^1$ continuum fluxes are in $10^{-16}~W~m^{-2}$, \\
$^2$ Luminosities in $10^{37}$ W, \\
$^3$ Fe line flux in $10^{-6}~photons~cm^{-2}~s^{-1}$)

\end{table}

\subsubsection{X-ray variability during ASCA observation}

\begin{figure}[t]
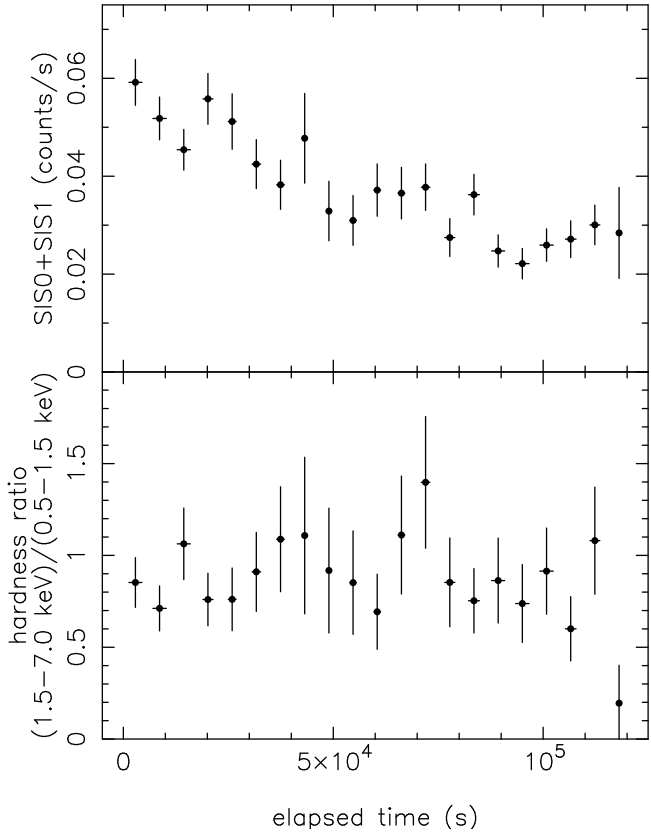

\clipfig{rxj0134_asca_light}{87}{10}{7}{200}{250}
\caption[ ]{\label{asca_light}
The ASCA light curve.  The upper panel shows the count rate variation.
The lower panel, showing the hardness ration as a function of time,
indicates that no significant spectral variability was detected.}
\end{figure}

\begin{figure}[t]
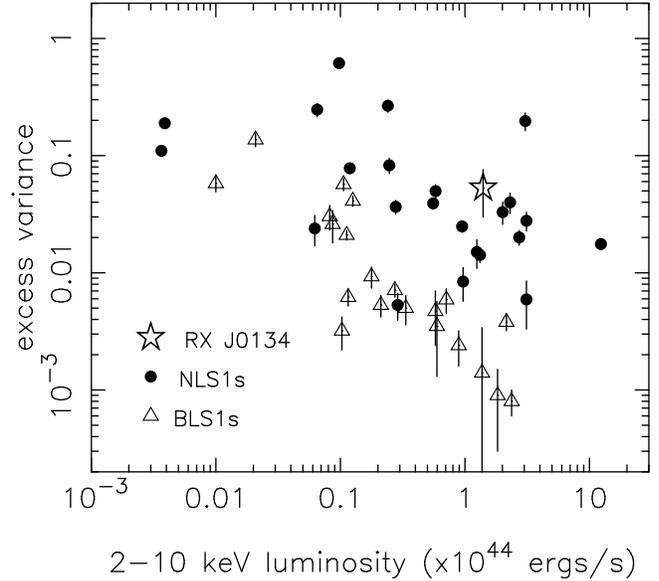

\clipfig{rxj0134_excess}{87}{14}{43}{200}{212}
\caption[ ]{\label{asca_excess}
Excess variance as function of X-ray luminosity.  RX J0134.2--4258 has
similar excess variance as other NLS1s. (Fig. adapted from Leighly
1999b).}

\end{figure}

We observed RX J0134.2--4258 to vary during the observation.  The
light curve from the SIS detectors binned by orbit is shown in Figure
\ref{asca_light}; a steady decline by  a factor of about 2 was
observed.  We calculated the fractional amplitude of variability, also
called the excess variance, defined as the measured variance minus the
measurement error and normalized by the square of the mean (e.g.
Nandra et al. 1997; Leighly 1999b).  This parameter is a useful one to
quantify variability in ASCA observations because it can be shown that
under a set of justifiable assumptions, it is related to the inverse
of the variability time scale (Leighly 1999b).  For a light curve with
128 second binning, the excess variance is $0.053 \pm 0.023$.  Figure
\ref{asca_excess}, adapted from Leighly (1999b), shows that this excess
variance is high compared with broad-line Seyfert galaxies with
similar luminosities but consistent with Narrow-line Seyfert 1
galaxies.  Therefore, although the hard X-ray spectrum appears to be
rather flat for an NLS1, the variability properties are unremarkable.  

\subsection{Optical}
Fig. \ref{opt_spec} displays the optical spectrum of RX
J0134.2--4258 obtained in September/October 1995.  The spectrum is
characteristic of a typical Narrow-line Seyfert 1 galaxy.  The broad
permitted blend of FeII emission commonly present in NLS1s severely
contaminates the spectrum, making an accurate measurement of the flux
and width, in particular of the H$\beta$ and O[III] lines difficult.
In order to account for this, we employ the FeII subtraction method
introduced by Boroson \& Green (1992) and now commonly used (e.g.\
Grupe et al.\ 1999; Leighly 1999a).  The FeII emission line spectrum
from a high signal-to-noise optical spectrum of the prototype strong
Fe emitter I Zw 1 is first convolved with a Gaussian and then scaled
until the width and intensity of the lines approximately match those
seen in the RX J0134.2--4258 spectrum.  This best-fitting FeII model
was then subtracted and the remaining emission lines examined.

We measured line widths of FWHM=900\pl100 and 670\pl200 $\rm
km~s^{-1}$ for H$\beta$ and [OIII]$\lambda$5007, respectively. The
ratio of the FeII to H$\beta$ emission is the strongest among all
objects in the entire sample of soft X-ray selected AGN ($\rm
\frac{FeII}{H\beta}$ = 12.3, see Grupe at al. 1999).  As we noted in
Grupe et al. 1998a, the object has a very blue spectrum ($\alpha_{\rm
opt}$ = --0.1).  We were not able to detect any [OIII]$\lambda$5007
emission in the spectra taken two years earlier; however, the [OIII]
is apparent in the better signal-to-noise spectrum presented here
(after FeII subtraction).  Otherwise, the results are the same as
obtained from our other spectra (Grupe et al. 1999).  

\newcommand{\putchartc}[1]{\clipfig{#1}{89}{10}{10}{280}{195}}
\newcommand{\chartlinec}[2]{\parbox[t]{18cm}{

\noindent\putchartc{#1}\charthsep\putchartc{#2}}}

\begin{figure*}[t]
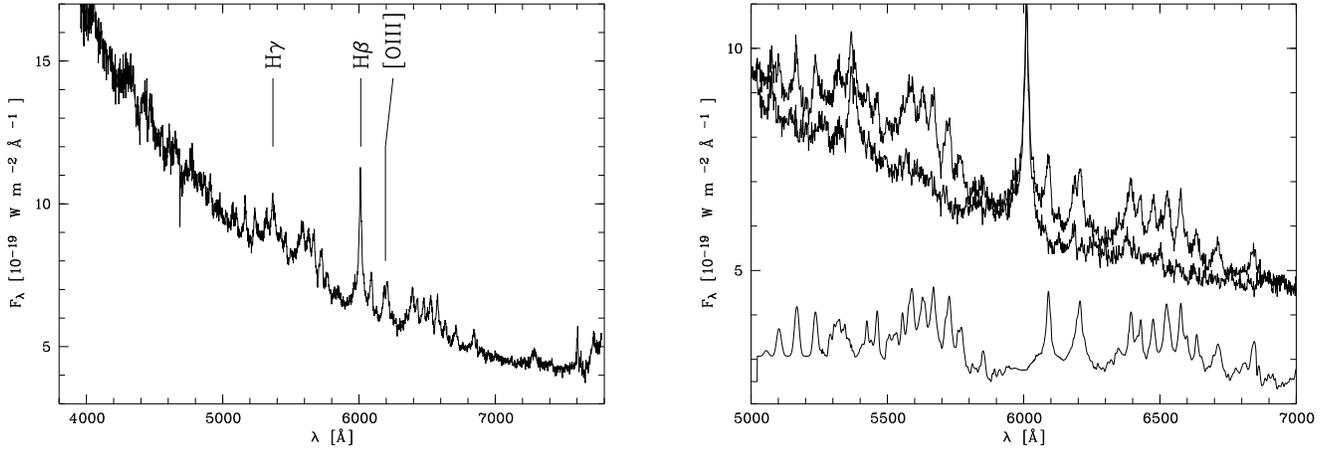

\chartlinec{rxj0134_opt}{rxj0134_av_fe2}
\caption[ ]{\label{opt_spec}
Optical spectrum of RX J0134.2--4258. The left panel shows the whole
wavelength range in which the object was observed, and the right panel
displays the FeII subtracted spectrum. The upper spectrum is the
original one, the middle one the FeII subtracted spectrum, the lower
one is the FeII template used, shown with an offset.
}
\end{figure*}

\subsection{Radio}

RX J0134.2--4258 was detected in the Parkes-MIT-NRAO (PMN) survey
(Wright et al.\ 1994) with a flux density of 55$\pm$9 mJy at 4.85 GHz.
Our new VLA observation yielded a flux of $25 \pm 1.5$ mJy at 8.4 GHz
and the source is completely unresolved in our map.  The two radio
fluxes allow us to derive a radio spectral index of $-1.4$, assuming
no variability between the two observations.  This radio spectral
index is fairly steep compared to core-dominated sources, but not
unheard of.  Applying the criterion of Kellermann et al. (1989), who
use R=10 as the dividing line between radio-loud and radio-quiet
quasars\footnote{R is defined as the K-corrected ratio of the 4.85 GHz
radio flux to the optical flux at 4400 A.}, the source is radio-loud,
since R=71.  The radio flux corresponds to a luminosity of log (P) =
25.3 W Hz$^{-1}$.  Thus, also using the luminosity criteria (e.g. Joly
1991; Miller et al.\ 1993), RX J0134.2--4258 should be classified as a
radio-loud AGN.  We note, however, that if the steep spectrum is real
and not a consequence of variability, the radio flux is highly
frequency dependent.

Narrow-line Seyfert 1 galaxies are generally considered to be weak
radio sources.  Ulvestad et al. (1995) found that the luminosities
among 15 NLS1s observed with the VLA were were generally less than log
P=22.5 W Hz$^{-1}$.  Only two other NLS1s are known to be radio-loud.
One is PKS~0558-504 (e.g. Remillard et al. 1991) and the other is
RGB~J0044+193 (Siebert et al. 1999).  Both of these objects are fairly
faint radio sources, with R $\approx 31$ and 27, respectively.  Thus,
RX~J0134.2--4258 has the relatively strongest radio emission of an
NLS1 so far discovered.

It is possible that RX~J0134.2--4258 is a variable radio source.  It
is not a catalogued member of the Parkes 2700 MHz survey catalog.  The
flux limit of this catalog depends on the location in the sky, but we
note that there are catalogued objects within 10 degrees with fluxes as
low as 50 mJy, although this appears to be the lower limit of the
catalog.  The PMN flux at 4.85 GHz predicts a flux of 125~mJy at 2700
MHz, assuming $\alpha=-1.4$; therefore, it should have been detected
by the Parkes survey unless it is variable.  Objects with steep
spectra are generally not variable, but since the observations
defining the index were not simultaneous, the steep index could in
fact be a consequence of variability.  Evidence for variability in the
radio has also been found in one of the other radio-loud NLS1s,
RGB~J0044+193 (Siebert et al. 1999).  On the other hand, many NLS1s
observed simultaneously at more than one radio frequency show steep
rather than flat spectra (Moran et al.\ in prep.).

\subsection{Spectral energy distribution}

\begin{figure}[t]
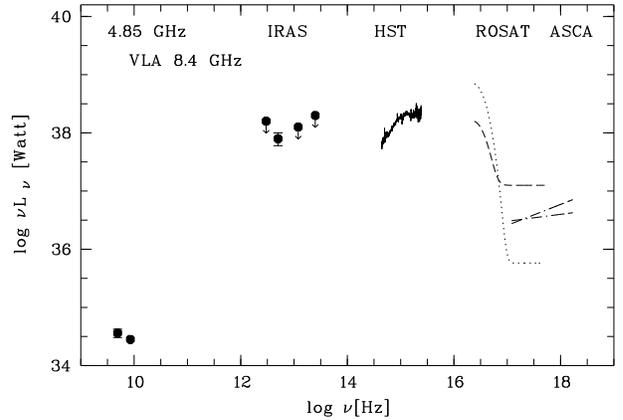

\clipfig{rxj0134_sed}{87}{10}{10}{280}{195}
\caption[ ]{\label{sed}
Spectral Energy Distribution 
of RX J0134.2--4258. The 4.85 GHz
radio point was taken from Wright et al. 1994, IRAS data
from Grupe et al. 1998a, and the optical/UV spectrum is from the HST data
(Goodrich et al., in prep). The dotted line represents a black body plus
power law spectrum of the RASS observation and the dashed line the same for the
pointed observation. The ASCA observation is represented by the dashed-dotted
line. All points are given in the observers frame.
}
\end{figure}

Fig. \ref{sed} displays the spectral energy distribution of our
source.  The object was also detected by IRAS at 60 $\mu$m and we
display upper limits at 12, 25, and 100 $\mu$m (Grupe et al. 1998a).
The ROSAT X-ray data in the plot are displayed as the black body plus
power law model fit to the RASS and pointed observation spectra. For
both cases, the radiation temperature and the hard X-ray spectral
slope were fixed (see Table \ref{fits}).

We used the nonsimultaneous {\it HST} (Goodrich et al., in prep) and
X-ray spectra to compute $\alpha_{ox}$, defined to be the
point-to-point slope between 2500\AA\ and 2~keV (Wilkes et al.\ 1994).
The pointed {\it ROSAT} observation, the {\it ASCA} observation and
the RASS observation yielded values of $\alpha_{ox}$ of 1.47, 1.68,
and 2.00 respectively.  Based on the regressions obtained by Wilkes et
al.\ 1994, the predicted values of $\alpha_{ox}$ are 1.61 and 1.43 for
radio-quiet and radio-loud AGN of this optical luminosity, respectively.  The
values obtained for RX~J0134.2--4258 lie between these predicted
indices, and in two cases are lower than the radio-quiet index.
Therefore, although it is impossible to account for the effect of
variability, the X-ray emission in RX~J0134.2--4258 does not appear to
be as strong as is found in other radio-loud AGN with similar optical
luminosities.

\section{\label{discussion} Discussion}

\subsection{Flux transient behaviour in ROSAT observations}

RX J0134.2--4258 was one of the softest AGN at the time of the RASS,
but turned into an object with a rather flat X-ray spectrum by the
time of its pointed observation.  Its behaviour is different from that
observed from other `transient' X-ray AGN.  WPVS007 was practically
`off' in all follow-up ROSAT observations after the RASS. We explained
this behaviour as a dramatic change in the comptonization parameters
that shifted the Big Blue Bump emission out of the ROSAT PSPC range
(Grupe et al. 1995b). In IC 3599 (and also NGC 5905) we observed a
fading of the X-ray flux over several years. We (Grupe et al.\ 1995a,
but see also Brandt at al.\ 1995; Bade, Komossa \& Dahlem 1996;
Komossa \& Bade 1999) came to the conclusion that these were X-ray
outbursts that could have been caused by tidal disruption of a star by
a central black hole.

RXJ0134.2--4258 can be considered to be a 'transient' X-ray source as
well.  During the RASS, it was practically off at energies above
$\sim$ 0.5 keV. While we usually find transient sources that were
bright during the RASS, such as WPVS007 or IC 3599, RX J0134--4258 is
an exception in that it became bright in its hard X-ray band after the
RASS.  What makes RX J0134.2--4258 so unusual is its dramatic spectral
variability, both on long time scales (between the RASS and pointed
observation) and on short time scales (during the pointed
observation).
 
\subsection{Precedents for similar behaviour}

The results from our study of RX~J0134.2--4258 are sufficiently
unusual that we have searched the literature for examples of 
similar behaviour from other objects.

The short term spectral variability observed in RX~J0134.2--4258 has
some precedent.  A hardness ratio analysis of the {\it ROSAT}
lightcurves from Mrk 766 revealed that the spectrum also hardens when
the flux increases.  Leighly et al.\ 1996 first demonstrated that this
behaviour was consistent with a non-varying soft excess component plus
a power law with varying normalization.  Those data have been recently
re-examined, and the same results were found (Page et al.\ 1999).

The long term spectral variability may also have some precedents.
Bedford, Vilhu \& Petrov (1988) report the discovery and followup {\it
EXOSAT} observations of the Seyfert 1.5 galaxy EXO~1128+691.  They
found the object clearly detected in the {\it EXOSAT} LE detector,
which is sensitive to soft photons in the 0.05--2.0~keV band, but not
detected by the ME detector (0.7--10 keV).  The LE detector has no
intrinsic energy resolution, but information from exposures using
several filters show that the spectrum was very soft, with effective
photon index $\Gamma \sim 5$. In contrast, a recent {\it ASCA}
observation of EXO~1128+691 showed a strong detection in 2--10 keV
X-rays (Leighly, Grupe, et al.\ in preparation).  The spectral
parameters from the {\it ASCA} observation predict 0.4 counts/s in the
{\it EXOSAT} ME.  While this count rate would be somewhat low for
detailed spectral analysis (e.g. Turner \& Pounds 1989), it should at
least have been detected.  Therefore, EXO~1128+691 may have undergone
a similar spectral transition as experienced by RX~J0134.2--4258.  A
difference is that during the {\it EXOSAT} observation, the LE flux
was strongly variable, by a factor of 3 on time scales as short as 15
minutes; in contrast, during the pointed {\it ROSAT} observation of
RX~J0134.2--4258 we find that the soft component is constant, and the
normalization of the hard component varies.

Another example of a possibly similar spectral transition is the one
recently seen in the Narrow-line Seyfert 1 galaxy NGC~4051 (Guainazzi
et al. 1998).  A {\it BeppoSAX} observation of this object found the
2--10 keV flux to be a factor of 20 lower than the historic average.
The remaining hard X-ray spectrum was very flat with a strong iron
line, suggesting a Compton-reflection dominated spectrum, as is found
in Compton-thick Seyfert 2 galaxies.  Guainazzi et al. (1998) report
that the soft X-ray spectrum was steep, but defer the details to
another paper.  While this also appears somewhat similar to the case
of RX~J0134.2--4258, a difference is that while the soft X-ray
component is reported to be relatively strong compared with the rest
of the spectrum, the {\it EUVE} light curve shows that it is much
weaker than has been seen in previous observations by {\it EUVE}
(J. Halpern 1999, P. comm.).

\subsection{Possible origins of the spectral variability}

\subsubsection{Spectral variability as a change in the warm absorber}

Komossa \& Fink (1997) and Komossa \& Meerschweinchen (2000) 
investigate whether the
spectral variability in RX J0134.2--4258 can be explained by a 'warm'
ionized absorber. She noted that there is a deviation in the residuals
of a power law fit to the pointed observation near 0.6keV (see Figure
\ref{po_spec}) that could be interpreted as an absorption edge from
OVII in the rest frame. In contrast, we found that a single power law
model does not give a good fit and the spectrum is much better
described using two component models in which case the residual
disappears.  This underlines the fact that with the spectral signal to
noise and the low resolution of the PSPC data, a two component model
and a warm absorber model cannot be distinguished.  The warm absorber
requires a very large column density ($10^{23}\rm cm^{-2}$) to explain
the ultrasoft spectrum, and we cannot accommodate such heavy
absorption in the {\it ASCA} spectra.  Furthermore, we do not see a
signature of this material in other wavebands.  For example, Reynolds
(1997) found that warm absorbers are generally associated with optical
reddening in a sample of {\it ASCA} observations of Seyfert 1
galaxies; in contrast, RX~J0134.2--4258 has one of the bluest optical
spectra observed in the soft X-ray selected AGN sample.

\subsubsection{Spectral variability as absence and recovery of  a corona}

One possible scenario to explain the longterm spectral variability
observed in RX J0134.2--4258 is the absence followed by the recovery
of the accretion disk corona. A corona is necessary to produce the
hard X-rays if the accretion disk is optically thick and geometrically
thin.  A popular version of the disk-corona model holds that the
accretion power is dissipated primarily in the corona, and the disk
merely reprocesses part of it (Haardt \& Maraschi 1993; Svensson \&
Zdziarski 1994).  However, it has recently been shown that in some
Narrow-line Seyfert 1 galaxies, there is far too much power in the
soft excess component, believed to be emitted from the disk, to be
driven by reprocessing of the hard X-ray component (e.g. Pounds, Done
\& Osborne 1995). Thus, the disk emission must be primary, the corona
will not be necessary as the primary agent for dissipation of the
accretion energy, and therefore it could conceivably be absent.  It is
possible that RX J0134.2--4258 had no corona during the RASS
observation, and it developed one before the pointed observation.  The
corona regeneration time scale must be fairly short for this scenario
to be viable.  If the corona is produced by reconnection of buoyant
magnetic flux tubes that have been built up to equipartition by
convective dynamos in the disk, the relevant time scales are the
convection time scales plus reconnection time scales (e.g. Haardt,
Maraschi \& Ghisellini 1994).  These authors find a loop variability
time scale could be as short as minutes for a $10^6 M_\odot$ object
and therefore the corona could conceivably regenerate within the time
between the RASS and pointed ROSAT observations.  The question of what
could cause an object to lose or regrow a corona is a very interesting
one that we will not speculate on.  On the other hand, it is not
necessarily clear that this mechanism could explain the spectral
variability observed on short time scales during the pointed
observation.

\subsubsection{Spectral variability as a strengthening of the radio component}

Another possible explanation for the spectral variability observed in
RX~J0134.2--4258 is suggested by the fact that it is radio-loud, and
possibly variable, and the fact that during the ASCA observation the
photon index was flat.  Radio-loud quasars are well known to have
flatter X-ray spectral indices than radio-quiet ones, and they are
also known to be brighter X-ray sources (e.g.  Lawson \& Turner 1997).
A widely accepted explanation for this fact is that the X-ray emission
includes a additional harder component associated with the radio jet.
Therefore, it is possible that the increase in the normalization of
the hard X-ray component is caused by a strengthening of the radio jet
component.  We note that compared with the {\it Ginga} sample of
quasars, RX~J0134.2--4258 is about a factor of 10 deficient in X-ray
luminosity (see Lawson \& Turner 1997, their Fig. 8).  However, taking into
account the scatter in the correlation and variability, it is not
completely implausible that the hard X-rays are being dominated by a
component associated with a radio jet.  

This scenario may have a
precedent in the case of 3C~273. A set of {\it ASCA} observations of
this object reveal that a soft excess and iron line appears, and the
spectrum becomes steeper when it is fainter (Cappi et al.\ 1998).  It
was proposed by Cappi et al. (1998) and also by Haardt et al. (1998)
that 3C~273 in its bright X-ray state is dominated by the jet
component, and in its faint state, the Seyfert nucleus is revealed.
On the other hand, it must be noted that {\it ASCA} observations of
the other two radio-loud NLS1s revealed canonical NLS1s spectra, i.e.
a steep hard X-ray spectrum and a weak soft excess (Siebert et al.
1999; Leighly 1999a).

The existence of radio-loud NLS1s may be somewhat surprising, given
the results of Ulvestad et al. (1995) on their small sample of
Seyferts, but also because NLS1s, and especially RX~J0134.2--4258, are
known to be strong emitters of Fe~II emission.  It has long been
considered that Fe~II emission is generally weak in radio-loud
objects, and [O III] is correspondingly strong.  Boroson \& Green
(1992) in their study of PG quasars found that one of the clearest
differences between radio-loud and radio-quiet objects is in their
Fe~II emission, although steep radio spectrum objects may have strong
Fe~II (also Joly 1991; Bergeron \& Kunth 1984).  However, the
situation could be similar to that for broad-absorption line quasars
(BALQSOs).  Early on it was found that BALQSOs are almost never
radio-loud (Stocke et al.\ 1992).  Later it was found that many
BALQSOs can be described as radio-moderate; they tend to lie at the
upper edge of the $R$ distribution for radio quiet quasars (Francis,
Hooper \& Impey 1993).  More recently it has been proposed that
radio-moderate or radio-intermediate objects are ones that are in fact
not radio-loud but rather have enhanced radio emission because they
have a weak radio jet which is beamed in our direction (Falcke,
Sherwood \& Patnaik 1996).  Such an interpretation may be attractive
for some NLS1s because of their observed rapid and possibly coherent
X-ray variability.  However, NLS1s do not appear to be universally
radio-intermediate.  RX~J0134.2--4258 is drawn from a soft X-ray
selected sample of AGN (e.g.\ Grupe 1996; Grupe et al.\ 1999).  A
search of the NRAO VLA Sky Survey (NVSS; Condon et al.\ 1993) reveals
that most of the northern ones are very weak radio emitters.

\section{Summary and conclusions}

We have presented the results of our ROSAT, ASCA and optical
observations of the enigmatic soft X-ray AGN RX J0134.2--4258. We
found:
 
\begin{itemize}
\item RX J0134.2--4258 has shown a dramatic change in its ROSAT PSPC spectra
between its RASS and a pointed PSPC observation made two years later.
\item Optically, RX J0134.2--4258 is a Narrow-Line Seyfert 1 galaxy with very
strong FeII emission, extremely weak emission from the NLR and with a very blue
optical continuum.
\item The slopes of the hard component of the ROSAT spectra and the ASCA 
spectra are consistent, and the ASCA slope is notably flat compared
with other NLS1s.
\item It is the normalization of the
hard X-ray component that varies the most and is primarily responsible for
the dramatic spectral change.
\item RX J0134.2--4258 has been found to be one of the few NLS1s that
are radio loud, and currently holds the record for radio-to-optical
flux ratio in NLS1s.
\item Physical reasons for the strange spectral behaviour might
be a variable warm absorber, the loss and regrowth of a corona above the
accretion disk, or an increase in flux from a component associated with a
possible jet as indicated by the radio-loudness of this source.
\end{itemize}

One important problem that applies to most transient AGN is that it is
difficult to monitor their behaviour. Often years lapsed between RASS
and later pointed observations. Hopefully, in the future with new
survey missions it will be easier and faster to discover transients.
Then they could be monitored using next-generation all sky monitor
experiments such as the approved mission {\it MAXI} which will be
mounted on the Japanese Experimental Module of the International Space
Station, or the proposed {\it Lobster-Eye} satellite.

\acknowledgements{We would like to thank 
Drs.~Beverley \& Derek Wills and Stefanie Komossa for useful and
intensive discussions.  We thank Dr.~S. Komossa for giving us a
preprint of her paper about NGC 5905 and her recent preprint on RX
J0134.2--4258.  Also we thank Drs. Thomas Boller, Joachim Siebert,
Jules Halpern and our referee
for their comments and suggestions on the manuscript.  
We wish
to thank Barry Clark and the NRAO staff for granting us 1 hour of
engineering time on short notice to obtain the new radio observations
which confirmed the association of the PMN radio source with RX
J0134.2--4258.  This research has made use of the NASA/IPAC
Extragalactic Database (NED) which is operated by the Jet Propulsion
Laboratory, Caltech, under contract with the National Aeronautics and
Space Administration. Also we used the IRAS data request of the
Infrared Processing and Analysis Center (IPAC) Caltech. 
We have made use of the ROSAT Data Archive of the Max-Planck-Institut 
f\"ur extraterrestrische Physik (MPE) at Garching, Germany. KML gratefully
acknowledges support through NAG5-7971 (NASA LTSA). SALM acknowledges
partial support by the Department of Energy at the Lawrence Livermore
National Laboratory under contract W-7405-ENG-48 and from the NSF
(grant AST-98-02791).  The ROSAT project is supported by the
Bundesministerium f\"ur Bildung und Forschung (BMBF/DLR) and the
Max-Planck-Society (MPG).

This paper can be retrieved via WWW from our pre-print server: \\
http://www.xray.mpe.mpg.de/~dgrupe/research/dgrupe.html}

   \end{document}

%% file: clipfig.tex
\def\clipfig#1{\def\lbracket{[}\def\testit{#1}%
    \ifx\testit\lbracket\let\next=\optclipfig\else\let\next=\stdclipfig\fi%
    \next{#1}}
%
\newcommand {\hclipfig} [7] {\clipfig[#7]{#1}{#2}{#3}{#4}{#5}{#6}}
%
\def\usemodepsfig {\global\def\cfmode{x}\typeout{*** set clipfig to PSFIG mode ***}}
\def\usemodeepsf  {\global\def\cfmode{}\typeout{*** set clipfig to EPSF mode ***}}
\def\useunitmm    {\global\def\cfunit{x}\typeout{*** set clipfig to use mm as unit ***}}
\def\useunitcm    {\global\def\cfunit{}\typeout{*** set clipfig to use cm as unit ***}}
\def\clipfigsettings {\ifx\cfmode\empty\def\ccfmode{EPSF }\else\def\ccfmode{PSFIG }\fi%
    \ifx\cfunit\empty\def\ccfunit{cm }\else\def\ccfunit{mm }\fi%
    \typeout{*** current clipfig settings: \ccfmode mode, using \ccfunit as unit ***}}
%
%
%
%
\def\stdclipfig#1#2#3#4#5#6{\ifx\cfmode\empty%
    \let\next=\eclipfig\else\let\next=\pclipfig\fi%
    \next{#1}{#2}{#3}{#4}{#5}{#6}}
\def\optclipfig#1#2]#3#4#5#6#7#8{\ifx\cfmode\empty%
    \let\next=\ehclipfig\else\let\next=\phclipfig\fi%
    \next{#3}{#4}{#5}{#6}{#7}{#8}{#2}}
%
%
%
\newcommand {\pclipfig}[6] {\ifx\cfunit\empty%
        \psfig{figure=#1.ps,width=#2cm,bbllx=#3cm,bblly=#4cm,bburx=#5cm,%
           bbury=#6cm,clip=}\else%
        \psfig{figure=#1.ps,width=#2mm,bbllx=#3mm,bblly=#4mm,bburx=#5mm,%
           bbury=#6mm,clip=}\fi}
\newcommand {\phclipfig}[7] {\ifx\cfunit\empty%
        \hspace{#7cm}\psfig{figure=#1.ps,width=#2cm,bbllx=#3cm,bblly=#4cm,%
           bburx=#5cm,bbury=#6cm,clip=}\else%
        \hspace{#7mm}\psfig{figure=#1.ps,width=#2mm,bbllx=#3mm,bblly=#4mm,%
           bburx=#5mm,bbury=#6mm,clip=}\fi}
%
%
%
\newcommand {\eclipfig}[6]{%
  \ifx\cfunit\empty\epsfxsize=#2cm\else\epsfxsize=#2mm\fi%
  \epsfclipon\epsfverbosetrue%
  \cfcmtopspts{#3}\cfllxi=\cftempi\cfllxf=\cftempf%
  \cfcmtopspts{#4}\cfllyi=\cftempi\cfllyf=\cftempf%
  \cfcmtopspts{#5}\cfurxi=\cftempi\cfurxf=\cftempf%
  \cfcmtopspts{#6}\cfuryi=\cftempi\cfuryf=\cftempf%
  \def\cfstra{\number\cfllxi.\number\cfllxf}%
  \def\cfstrb{\number\cfllyi.\number\cfllyf}%
  \def\cfstrc{\number\cfurxi.\number\cfurxf}%
  \def\cfstrd{\number\cfuryi.\number\cfuryf}%
  \hbox{\epsfbox[{\cfstra} {\cfstrb} {\cfstrc} {\cfstrd}]{#1.ps}}}
\newcommand {\ehclipfig}[7]{%
  \ifx\cfunit\empty\epsfxsize=#2cm\else\epsfxsize=#2mm\fi%
  \epsfclipon\epsfverbosetrue%
  \cfcmtopspts{#3}\cfllxi=\cftempi\cfllxf=\cftempf%
  \cfcmtopspts{#4}\cfllyi=\cftempi\cfllyf=\cftempf%
  \cfcmtopspts{#5}\cfurxi=\cftempi\cfurxf=\cftempf%
  \cfcmtopspts{#6}\cfuryi=\cftempi\cfuryf=\cftempf%
  \def\cfstra{\number\cfllxi.\number\cfllxf}%
  \def\cfstrb{\number\cfllyi.\number\cfllyf}%
  \def\cfstrc{\number\cfurxi.\number\cfurxf}%
  \def\cfstrd{\number\cfuryi.\number\cfuryf}%
  \ifx\cfunit\empty\hspace{#7cm}\else\hspace{#7mm}\fi%
  \hbox{\epsfbox[{\cfstra} {\cfstrb} {\cfstrc} {\cfstrd}]{#1.ps}}%
  \vspace{-1mm}}
%
%
%
\newdimen\cfllxi \newdimen\cfllyi  \newdimen\cfurxi  \newdimen\cfuryi
\newdimen\cfllxf \newdimen\cfllyf  \newdimen\cfurxf  \newdimen\cfuryf
\newdimen\cftemp \newdimen\cftempi \newdimen\cftempf
\newdimen\cfpspoint \cfpspoint=1bp
%
%
%
\newcommand{\cfcmtopspts}[1]{\ifx\cfunit\empty%
  \cftemp=#1cm\else\cftemp=#1mm\fi%
  \multiply\cftemp10\divide\cftemp\cfpspoint%
  \cftempf=\cftemp\divide\cftemp10\cftempi=\cftemp\multiply\cftemp10%
  \advance\cftempf-\cftemp}
%
%
\def\cfmode{}\def\cfunit{}\clipfigsettings
%

%% file: rxj0134.bbl
\begin{thebibliography}{}
\bibitem{ } Bade N., Komossa S., Dahlem M., 1996, A\&A 309, L35
\bibitem { } Bedford D.K., Vilhu O., Petrov P., 1988, MNRAS 234, 319
\bibitem { } Bergeron J., Kunth D., 1984, MNRAS 207, 263
\bibitem { } Beuermann K., Thomas H.-C., Reinsch K., et  al., 1999, A\&A 347, 47
\bibitem[1992]{bg} Boroson T.A., Green R.F., 1992, ApJS 80, 109 
\bibitem[1995]{bra} Brandt W.N., Pounds K.A., Fink H.H., 1995, MNRAS 273, L47
\bibitem { } Cappi M., Matsuoka M., Otani C., Leighly K.M., 1998, PASJ 50, 213
\bibitem{ } Condon, J. J., Cotton, W. D., Greisen, E. W., Yin, Q. F.,
Perley, R. A., Taylor, G. B., \& Broderick, J. J., 1998, AJ, 115, 1693
\bibitem[1990]{dl} Dickey J.M., Lockman F.J., 1990, ARA\&Astrophys. 28, 215
\bibitem { } Falcke, H., Sherwood, W., \& Patnaik, A. R., 1996, ApJ,
471, 106
\bibitem { } Francis, P. J., Hooper, E. J., \& Impey, C. D., 1993, AJ,
106, 417
\bibitem { } Guainazzi M., Nicastro F., Fiore F., et al., 1998, MNRAS 301, L1
\bibitem[1995c]{grupe3} Grupe, D., Beuermann, K., Mannheim, K.,
et al., 1995a, A\&A 299, L5
\bibitem[1995d]{grupe4} Grupe D., Beuermann K., Mannheim K., et al., 1995b,
A\&A 300, L21
\bibitem{ } Grupe D., 1996, ``{\it Properties of Bright Soft X-ray Selected
ROSAT AGN}'', PhD Thesis, Universit\"at G\"ottingen 
(available via our pre-print server)
\bibitem[1995a]{grupe1} Grupe D., Beuermann K., Mannheim K., Thomas H.-C.,
Fink, H.H., 1998a, A\&A 330, 25 
\bibitem[ ]{1997} Grupe D., Wills B.J., Wills D., Beuermann, K., 1998b,
A\&A 333, 827
\bibitem { } Grupe D., Beuermann K., Mannheim K., Thomas H.-C., 1999, A\&A
350, 805
\bibitem { } Haardt F., Maraschi L., 1993, ApJ 413, 507
\bibitem { } Haardt F., Maraschi L., Ghisellini G., 1994, ApJL 432, 95
\bibitem { } Haardt F., Fossati G., Garndi P., et al., 1998, A\&A 340, 35
\bibitem { } Iwasawa K., Fabian A.C., Nandra K., 1999, MNRAS, in press
\bibitem { } Joly M., 1991, A\&A 242, 49
\bibitem{ } Kellermann K.I., Sramek R., Schmidt M., Shaffer D.B., Green R., 
1989, AJ 98, 1195
\bibitem { } Komossa S., Fink H.H., 1997, ``Accretion Disks - New Aspects'',
Lecture Notes in Physics 487, 250
\bibitem { } Komossa S., Bade N., 1999, A\&A 343, 775
\bibitem { } Komossa S., Meerschweinchen J., 2000, A\&A in press
\bibitem { } Lawson A.J., Turner M.J.L., 1999, MNRAS 288, 920
\bibitem { } Leighly K.M., Mushotzky R.F., Yaqoob T., Kunieda H., Edelson, R.,
1996, ApJ 469, 147
\bibitem{ } Leighly K.M., 1999a, ApJS in press
\bibitem{ } Leighly K.M., 1999b, ApJS in press
\bibitem[1995]{man}
Mannheim K., Grupe D., Beuermann K., Thomas H.-C., Fink H.H., 1996,
MPE Report 263, ``R\"ontgenstrahlung from the Universe'', p471
\bibitem{ } Miller P., Rawlings S., Saunders R., 1993, MNRAS 263, 425
\bibitem{ } Nandra K., George I. M., Mushotzky R. F., Turner T. J.,  Yaqoob,
T., 1997, ApJ 476, 70
\bibitem { } Page M.J., Carrera F.J., Mittaz J.P.D., Mason K.O., 1999, MNRAS in
press
\bibitem { } Pfeffermann E., Briel U.G., Hippmann H., et al., 1986, 
SPIE 733, 519
\bibitem { } Pounds K.A., Done C., Osborne J., 1996, MNRAS 277, L5
\bibitem { } Remillard R.A., Grossan B., Bradt H.V., Ohashi T., Hayashida K., 
1991, Nature 350, 589
\bibitem { } Reynolds C.S., 1997, MNRAS 286, 513
\bibitem { } Siebert J., Leighly K.M., Laurent-Muehleisen S.A., et al., 1999,
A\&A 348, 678
\bibitem { } Stocke, J. T., Morris, S. L., Weymann, R. J., \& Foltz,
C. B., 1992, ApJ, 396, 487
\bibitem { } Svensson R., Zdziarski A.A., 1994, ApJ 436, 599
\bibitem { } Tanaka Y., Inoune H., Holt S.S., 1994, PASJ 46, L37
\bibitem{hct} Thomas H.-C., Beuermann K., Reinsch K., et al., 1998,
A\&A 335, 467
\bibitem{true} Tr\"umper J., 1983, Adv. Space Res. 4, 241
\bibitem { } Turner T.J., Pounds K.A., 1989, MNRAS 240, 833
\bibitem { } Ulvestad J.S., Antonucci, R.R.J., Goodrich, W., 1995, AJ 109, 81
\bibitem{voges} Voges W., 1993, Adv. Space. Res. 13, 391
\bibitem{ } Voges W., Aschenbach B., Boller Th., et al., 1999, A\&A 349, 389
\bibitem { } Wilkes, B. J., Tananbaum, H., Worrall, D. M., Avni, Y.,
Oey, M. S., \& Flanagan, J., 1994, ApJS, 92, 53
\bibitem[ ]{ } Wright A.E.,  Griffith M.R., Burke B.F., Ekers R.D.,
1994, ApJS 91, 111 
\bibitem[1990]{ } Zimmermann, H.U., G. Boese, G., W. Becker, W., et al.,
1998, EXSAS User's Guide, MPE Report
\end{thebibliography}
